\documentclass[prx,superscriptaddress,onecolumn,showpacs]{revtex4-1}
\usepackage{hyperref}
\usepackage{graphicx,epsfig}
\usepackage{subeqn}
\usepackage[usenames]{color}

\newcommand{\bk}{{\bf k}}
\newcommand{\bl}{{\bf l}}
\newcommand{\mm}{{\bf m}}

\newcommand{\cE}{{\mathcal E}}
\newcommand{\cG}{{\mathcal G}}
\newcommand{\cH}{{\mathcal H}}

\newcommand{\cP}{{\mathcal P}}
\newcommand{\cS}{{\mathcal S}}

\newcommand{\cZ}{{\mathcal Z}}

\begin{document}


\title{The role of the chemical potential in the BCS theory}

\date{\today}

\author{Drago\c s-Victor Anghel}
\email{dragos@theory.nipne.ro}
\affiliation{Horia Hulubei National Institute for Physics and Nuclear Engineering, P.O. Box MG-6, 077126 M\u agurele, Ilfov, Romania}

\author{George Alexandru Nemnes}
\affiliation{Horia Hulubei National Institute for Physics and Nuclear Engineering, P.O. Box MG-6, 077126 M\u agurele, Ilfov, Romania}
\affiliation{University of Bucharest, Faculty of Physics, ``Materials and Devices for Electronics and Optoelectronics'' Research Center, 077125 M\u agurele-Ilfov, Romania}

\begin{abstract}
We study the effect of the chemical potential on the results of the BCS theory of superconductivity.
We assume that the pairing interaction is manifested between electrons of single-particle energies in an interval $[\mu - \hbar\omega_c, \mu + \hbar\omega_c]$, where $\mu$ and $\omega_c$ are parameters of the model--$\mu$ needs not be equal to the chemical potential of the system, denoted here by $\mu_R$. The BCS results are recovered if $\mu = \mu_R$. If $\mu \ne \mu_R$ the physical properties change significantly: the energy gap $\Delta$ is smaller than the BCS gap, a population imbalance appears, and the superconductor-normal metal phase transition is of the first order.
The quasiparticle imbalance is an equilibrium property that appears due to the asymmetry with respect to $\mu_R$ of the single-particle energy interval in which the pairing potential is manifested.

For $\mu_R - \mu$ taking values in some ranges, the equation for $\Delta$ may have more than one solution at the same temperature, forming branches of solutions when $\Delta$ is plotted vs $\mu_R-\mu$ at fixed $T$.
The solution with the highest energy gap, which corresponds to the BCS solution when $\mu = \mu_R$, cease to exist if $|\mu-\mu_R| \ge 2\Delta_0$ ($\Delta_0$ is the BCS gap at zero temperature).
Therefore the superconductivity is conditioned by the existence of the pairing interaction and also by the value of $\mu_R - \mu$. \\

Theories and models of superconducting state
BCS theory and its development
Effects of pressure
Other effects in superconductivity

\end{abstract}

\maketitle

\section{Introduction} \label{intro}

In the BCS formalism of superconductivity \cite{Tinkham:book,PhysRev.108.1175.1957.Bardeen}, an attractive interaction potential is assumed to act between pairs of electrons with opposite momenta and opposite spin projections (or between pairs of electrons of time reversed eigenstates \cite{JPhysChemSolids.11.26.1959.Anderson}) if their free single-particle energies $\epsilon_\bk^{(0)}$ are in an interval of width $2\hbar\omega_c$ centered at 
$\mu$.
In general it is assumed that $\mu$ is one and the same as the chemical potential of the system, which will be denoted here by $\mu_R$ and which is determined by the density of electrons and the interaction of the system with the environment (e.g. with a heat and particle reservoir).
We consider that the pairing potential is a microscopic characteristic of the interaction between the electrons and therefore it is not (or at least not directly) determined by the value of the chemical potential, which may be changed by applied pressure or doping. 
For this reason, we shall assume that $\mu_R$ may take values which are different from $\mu$ and we shall investigate the effects of this difference on the superconducting phase.

As expected, when $\mu = \mu_R$ we recover the typical BCS theory.
If $\mu\ne\mu_R$ the thermodynamics change significantly, namely the gap energy $\Delta$ and the temperature of the superconductor-normal metal phase transition decreases with $|\mu-\mu_R|$, a quasiparticle branch imbalance appears, and the phase transition becomes discontinuous (of the first order).
For some intervals of $\mu_R-\mu$, the equation for $\Delta$ may have multiple solutions at the same temperature. For this reason the plot $\Delta$ vs $\mu_R-\mu$ at fixed $T$ shows multiple branches.
In this paper we analyze only the branch of solutions with the biggest gap, which reduce to the BCS solution in the point $\mu_R = \mu$.
For this branch of solutions, if $|\mu_R - \mu| < 2\Delta_0$ ($\Delta_0$ is the BCS energy gap at zero temperature) the zero temperature gap is independent of $\mu-\mu_R$ (although the temperature of the phase transition decreases to zero as $|\mu-\mu_R|$ increases to $2\Delta_0$), whereas for $|\mu-\mu_R| > 2\Delta_0$ the energy gap is not formed anymore, even at zero temperature.
In the standard BCS theory, this phenomenon may be the equivalent of the \textit{superconducting dome} from the high-$T_c$ superconductivity.

It is known that quasiparticle imbalance appears when quasiparticle currents are injected into the superconductor sufficiently high above the energy gap. This is a non-equilibrium situation and is explained in Refs. \cite{PhysRevLett.28.1363.1972.Clarke, PhysRevLett.28.1366.1972.Tinkham, PhysRevB.6.1747.1972.Tinkham} by attributing different chemical potentials to the quasiparticle system and to the condensate.
We show here that quasiparticle imbalance appears also in equilibrium, due to the asymmetry with respect to $\mu_R$ of the interval in which the pairing interaction is manifested and when there is no distinction between the chemical potential of the pairs and the chemical potential of the quasiparticles (related to this, see the discussion in Ref. \cite{PhysRevB.92.144506.2015.Miller}).

The paper is organized as follows. In the next subsection we present the basic results of the BCS theory and the Bogoliubov transformations, in order to introduce the notations and to emphasize that we use the standard formalism. In Sections \ref{sec_N} and \ref{sec_E} we calculate the total number of particles $N$ and the energy $E$ of the system, whereas in the Sections \ref{Var_en_gap} and \ref{Var_num_part} we calculate the variations of $E$ and $N$ with the populations of the quasiparticle states. In Section \ref{sec_Eq_ples} we calculate the equilibrium particle distribution and present the main results. The conclusions are presented in Section \ref{discussion}.

Our method of calculation resembles the one presented in Ref.~\cite{PhysicaA.459.276.2016.Anghel}, with the difference that here we introduce the additional parameter $\mu_R$ and we calculate the dependence of the average number of particles $N$ on the number of quasiparticle excitations, which is then used in the maximization of the partition function.
The average number of particles in the system is determined by the populations of the quasiparticle states and by the parameters of the model Hamiltonian, $\mu$, $\omega_c$, and $V$.
This more rigorous approach introduces an extra term in the quasiparticle excitation energy, which determines the properties of the system when $\mu \ne \mu_R$, but has no effect on the equilibrium properties when $\mu \equiv \mu_R$ and constant density of states, which is the simplified situation discussed in Ref.~\cite{PhysicaA.459.276.2016.Anghel}.

%

\subsection{Standard BCS approach} \label{subsec_BCS}


Let us consider the BCS formalism for clean superconductors, in which the electrons' wavefunctions are plane waves of wavevectors $\bk$. If we denote by $s=\pm 1/2$ (or $s=\uparrow,\downarrow$) the spin projection, then the state of the electron is denoted by $|\bk,s\rangle$.
We denote the electrons creation and annihilation operators by $c^\dagger_{\bk, s}$ and $c_{\bk,s}$, respectively, and the BCS pairing Hamiltonian is \cite{Tinkham:book,PhysRev.108.1175.1957.Bardeen}
\begin{equation}
  \hat\cH_{BCS} = \sum_{\bk s}\epsilon^{(0)}_\bk n_{\bk s} + \sum_{\bk\bl} V_{\bk\bl} c^\dagger_{\bk\uparrow} c^\dagger_{-\bk\downarrow} c_{-\bl\downarrow} c_{\bl\uparrow} , \label{def_H_BCS}
\end{equation}
where $n_{\bk s} \equiv c^\dagger_{\bk, s} c_{\bk,s}$ is the occupation number operator for the single-particle state $|\bk,s\rangle$ and $\epsilon_\bk^{(0)}$ is the corresponding single-particle energy.
Following for example the textbook \cite{Tinkham:book}, to diagonalize the Hamiltonian (\ref{def_H_BCS}) one introduces the notation $b_\bk = \langle c_{-\bk\downarrow} c_{\bk\uparrow} \rangle$ (where $\langle \cdot \rangle$ is the average) and writes $c_{-\bk\downarrow} c_{\bk\uparrow} \equiv b_\bk + (c_{-\bk\downarrow} c_{\bk\uparrow} - b_\bk)$.
Writing $\cH_{BCS}$ in terms of this difference and keeping only the first order terms we get
\begin{equation}
  \hat\cH = \sum_{\bk s}\epsilon^{(0)}_\bk n_{\bk s} + \sum_{\bk\bl} V_{\bk\bl} ( c^\dagger_{\bk\uparrow} c^\dagger_{-\bk\downarrow} b_\bl + b^*_\bk c_{-\bl\downarrow} c_{\bl\uparrow} - b^*_\bk b_\bl). \label{def_H_BCS_approx}
\end{equation}
The Hamiltonian (\ref{def_H_BCS_approx}) is quadratic and can be diagonalized.
To obtain analytic expressions and convergent quantities in the thermodynamic limit (convergent energy gap, condensation energy, etc.), one assumes that the pairing potential $V_{\bk\bl} \equiv V$ is constant and different from zero only if $\epsilon_\bk^{(0)}$ and $\epsilon_\bl^{(0)}$ belong to a finite interval $I_V \equiv [\mu - \hbar\omega_c, \mu + \hbar\omega_c]$ centered at the \textit{BCS chemical potential} $\mu$.
For the convenience of the calculations, $\hat\cH$ is transformed into the model Hamiltonian $\hat\cH_M = \hat\cH - \mu \hat N$, where $\hat N \equiv \sum_{\bk,s} c^\dagger_{\bk s} c_{\bk s}$ is the particle number operator. $\hat\cH_M$ is diagonalized to become
\begin{equation}
  \hat\cH_M = \sum_\bk(\xi_\bk-\epsilon_\bk+\Delta b_\bk^*) + \sum_\bk\epsilon_\bk(\gamma^\dagger_{\bk 0}\gamma_{\bk 0} + \gamma^\dagger_{\bk 1}\gamma_{\bk 1}) , \label{HM_BCS}
\end{equation}
where $\xi_\bk\equiv \epsilon^{(0)}_\bk - \mu$, $\epsilon_\bk \equiv \sqrt{\xi_\bk^2+\Delta^2}$, and $\Delta$ is the \textit{energy gap}, which is defined by the equation
\begin{eqnarray}
  \Delta &=& - V \sum_\bl \langle c_{-\bk\downarrow} c_{\bk\uparrow}\rangle \label{def_Delta0}
\end{eqnarray}
The operators $\gamma^\dagger_{\bk i}$ and $\gamma_{\bk i}$ ($i=0,1$) are the quasiparticle creation and annihilation operators, respectively, and are defined by the relations
$  c_{\bk\uparrow} = u^*_\bk \gamma_{\bk 0} + v_\bk \gamma^\dagger_{\bk 1} , \ 
  c^\dagger_{\bk\uparrow} = u_\bk \gamma^\dagger_{\bk 0} + v^*_\bk \gamma_{\bk 1} , \ 
  c^\dagger_{-\bk\downarrow} = -v^*_\bk \gamma_{\bk 0} + u_\bk \gamma^\dagger_{\bk 1} , \ {\rm and}\ 
  c_{-\bk\downarrow} = -v_\bk \gamma^\dagger_{\bk 0} + u^*_\bk \gamma_{\bk 1}$,
where $u_\bk$ and $v_\bk$ are
\begin{equation}
  |v_\bk|^2 = 1 - |u_\bk|^2 = \frac{1}{2} \left(1 - \frac{\xi_\bk}{\epsilon_\bk}\right) . \label{def_uv}
\end{equation}
Using the definitions of $\Delta$ and of the $\gamma$ operators, Eq. (\ref{def_Delta0}) is transformed into the self-consistent equation for the gap energy
\begin{equation}
  1 = \frac{V}{2} \sum_\bl \frac{1 - n_{\bk 0} - n_{\bk 1}}{\epsilon_l} , \label{def_Delta2}
\end{equation}
where $n_{\bk i} = \gamma^\dagger_{\bk i}\gamma_{\bk i}$ is the number of quasiparticles.
If we work in the quasicontinuous limit, the summation over $\bk$ is transformed into an integral over the free particle energies $\epsilon^{(0)}$ with the general density of states (DOS) $\sigma(\epsilon^{(0)}) \equiv \sigma(\xi+\mu)$.
Then Eq. (\ref{def_Delta2}) becomes
\begin{eqnarray}
  \frac{2}{V} &=& \int_{-\hbar\omega_c}^{\hbar\omega_c} \frac{1 - n_{0\epsilon^{(0)}} - n_{1\epsilon^{(0)}}}{\sqrt{\xi^2+\Delta^2}} \sigma(\xi+\mu)\,d\xi .
  \label{Eq_int_Delta1}
\end{eqnarray}
The equation for the gap energy at zero temperature $\Delta_0 \equiv \Delta(T=0)$ is
\begin{eqnarray}
  \frac{2}{V} &=& \int_{-\hbar\omega_c}^{\hbar\omega_c} \frac{\sigma(\xi+\mu)\,d\xi}{\sqrt{\xi^2+\Delta_0^2}} . \label{Eq_int_Delta2}
\end{eqnarray}
If the DOS is constant, i.e. $\sigma(\xi+\mu) \equiv\sigma_0$, Eq. (\ref{Eq_int_Delta2}) gives the usual BCS result,
$\Delta_0 = 2\hbar\omega_c\exp[-1/(\sigma_0V)]$.

\section{The number of particles in the system} \label{sec_N}

The only BCS parameters that we specified in Section \ref{subsec_BCS} are $\mu$, $\omega_c$, and $V$.
We denote the BCS ground state (the state with no quasiparticle excitations) by
\begin{equation}
  |\{0\},\mu\rangle = \prod_{\bk} (u_\bk +v_\bk c^\dagger_{\bk\uparrow}c^\dagger_{-\bk\downarrow}) |0\rangle , \label{def_BCS_gs}
\end{equation}
where $u_\bk$ and $v_\bk$ were defined in Section \ref{subsec_BCS} and $|0\rangle$ is the vacuum state.
A quasiparticle creation operator applied to the BCS state creates a single-particle excitation,
\begin{subequations}\label{def_BCS_sp_excs}
\begin{eqnarray}
  \gamma^\dagger_{\bl 0}|\{0\},\mu\rangle &=& c^\dagger_{\bl \uparrow} \prod_{\bk(\ne\bl)} (u_\bk +v_\bk c^\dagger_{\bk\uparrow}c^\dagger_{-\bk\downarrow}) |0\rangle , \label{def_BCS_sp_exc0} \\
  \gamma^\dagger_{\bl 1}|\{0\},\mu\rangle &=& c^\dagger_{-\bl \downarrow} \prod_{\bk(\ne\bl)} (u_\bk +v_\bk c^\dagger_{\bk\uparrow}c^\dagger_{-\bk\downarrow}) |0\rangle, \label{def_BCS_sp_exc1}
\end{eqnarray}
whereas two gamma operators applied to the same pair of states leads to a pair excitation,
\begin{eqnarray}
  \gamma^\dagger_{\bl 0}\gamma^\dagger_{\bl 1}|\{0\},\mu\rangle &=& (u_\bk c^\dagger_{\bl \uparrow} c^\dagger_{-\bl \downarrow} - v_\bl) \prod_{\bk(\ne\bl)} (u_\bk +v_\bk c^\dagger_{\bk\uparrow}c^\dagger_{-\bk\downarrow}) |0\rangle = -\gamma^\dagger_{\bl 1}\gamma^\dagger_{\bl 0}|\{0\},\mu\rangle. \label{def_BCS_p_exc21}
\end{eqnarray}
\end{subequations}
In the presence of the set of excitations denoted by $\{n_{\bk i}\}$ ($i=0,1$), the state of the system is denoted by $|\{n_{\bk i}\},\mu\rangle$ and $\Delta$ is determined from Eq. (\ref{def_Delta2}) or (\ref{Eq_int_Delta1}).
Following Ref. \cite{PhysRev.108.1175.1957.Bardeen}, we denote by $\cG$ the set of pairs in the ground state, by $\cP$ the set of excited pairs (Eq. \ref{def_BCS_p_exc21}), and by $\cS$ the set of single-particle excitations (Eqs. \ref{def_BCS_sp_exc0} and \ref{def_BCS_sp_exc1}).
Then the expectation value for the particle number operator is
\begin{eqnarray}
  N &\equiv& \langle\{n_{\bk i}\},\mu|\hat N|\{n_{\bk i}\},\mu\rangle
  = N' + \sum_{\bk\in \cG}2v^2_{\bk} + \sum_{\bl\in \cP} 2u^2_{\bl} + \sum_{\mm\in \cS} 1
  %
  \equiv N_0 + \sum_{\bk, i} n_\bk \frac{\xi_\bk}{\epsilon_\bk}
  %
  , \label{N_exp_val1}
\end{eqnarray}
where $N'$ is the contribution coming from outside of the energy interval $[\mu-\hbar\omega_c,\mu+\hbar\omega_c]$ (not explicitly written here), whereas $N_0 = N' + \sum_{\bk}2v^2_{\bk}$ is the number of particles without taking into account the contribution of the excitations (although the excitations determine $\Delta$ and therefore $v_\bk$, as we shall see below).
We observe that each excitation contributes to $N-N_0$ by $\delta N_\xi = \xi/\epsilon$, with no difference between single-particle or pair excitations.

If we denote by $N_\mu \equiv 2 \sum_{\bk}^{k\le k_F} 1$ the number of free single-particle states up to the level $\epsilon^{(0)} = \mu$ ($N_\mu$ is a constant), then
\begin{eqnarray}
  N &=& N_\mu - \sum_{\xi = -\hbar\omega_c}^{0}\left[ 1 - \frac{|\xi|}{\epsilon} + \left( n_{\xi 0} + n_{\xi 1} \right) \frac{|\xi|}{\epsilon} \right]
  + \sum_{\xi = 0}^{\hbar\omega_c}\left[ 1 - \frac{\xi}{\epsilon} + \left( n_{\xi 0} + n_{\xi 1} \right) \frac{\xi}{\epsilon} \right] . \label{N_Nmu_sum}
\end{eqnarray}
Equation (\ref{N_Nmu_sum}) in the continuous limit becomes
\begin{eqnarray}
  \langle N \rangle - N_\mu 
  %
  &=& \int_{0}^{\hbar\omega_c}[\sigma(\xi+\mu) - \sigma(-\xi+\mu)] \left( 1 - \frac{\xi}{\epsilon} \right) d\xi
  + \int_{-\hbar\omega_c}^{\hbar\omega_c}\sigma(\xi+\mu) ( n_{\xi 0} + n_{\xi 1}) \frac{\xi}{\epsilon} d\xi
  \label{N_Nmu_int}
\end{eqnarray}
If the DOS is constant, then
\begin{eqnarray}
  \langle N \rangle - N_\mu 
  %
  &=& \sigma_0 \int_{-\hbar\omega_c}^{\hbar\omega_c} ( n_{\xi 0} + n_{\xi 1}) \frac{\xi}{\epsilon} d\xi . \label{N_Nmu_int_sconst}
\end{eqnarray}
We shall see that in equilibrium $n_{\xi 0} = n_{\xi 1} \equiv n_\xi$ for any $\xi$, so Eq. (\ref{N_Nmu_int_sconst}) becomes
\begin{eqnarray}
  \langle N \rangle &=& N_\mu + 2 \sigma_0 \int_{0}^{\hbar\omega_c} ( n_{\xi} - n_{-\xi}) \frac{\xi}{\epsilon} d\xi
   = 2 \sigma_0 \int_{\Delta}^{\hbar\omega_c} ( n_{\sqrt{\epsilon^2 - \Delta^2}} - n_{-\sqrt{\epsilon^2 - \Delta^2}}) d\epsilon
  \label{N_Nmu_int_sconst2}
\end{eqnarray}

\section{The energy of the system} \label{sec_E}

The energy of the system may be written as
\begin{equation}
  \cH = E_0 + \sum_\bk\epsilon_\bk (\gamma^\dagger_{\bk 0}\gamma_{\bk 0} + \gamma^\dagger_{\bk 1}\gamma_{\bk 1}) , \label{HM_BCS_eff}
\end{equation}
where
\begin{equation}
  E_0 = \mu N + \sum_\bk(\xi_\bk-\epsilon_\bk+\Delta b_\bk^*)
  \equiv \mu N + \sum_\bk (\xi_\bk - \epsilon_\bk) + \frac{\Delta^2}{V} . \label{def_E0}
\end{equation}
%
%
We denote by $E_{0\mu}$ the energy of the free electron gas with the number of particles equal to $N_\mu$ and at zero temperature ($E_{0\mu}$ is a constant). Using Eq. (\ref{N_Nmu_int}) we calculate
\begin{eqnarray}
  E_0 - E_{0\mu} &=& \sum_{k> k_\mu} (\xi_\bk - \epsilon_\bk) + \sum_{k \le k_\mu} (- \xi_\bk - \epsilon_\bk) + \frac{\Delta^2}{V} + \mu (\langle N \rangle - N_\mu) \nonumber \\
  %
  %
  &=& \frac{\Delta^2}{V} - \int_{-\hbar\omega_c}^{\hbar\omega_c} \sigma(\xi+\mu) \epsilon_\xi \left(1 - \frac{|\xi|}{\epsilon} \right)\,d\xi + \mu \int_{0}^{\hbar\omega_c}[\sigma(\xi+\mu) - \sigma(-\xi+\mu)] \left( 1 - \frac{\xi}{\epsilon} \right) d\xi \nonumber \\
  && + \mu \int_{-\hbar\omega_c}^{\hbar\omega_c}\sigma(\xi+\mu) ( n_{\xi 0} + n_{\xi 1}) \frac{\xi}{\epsilon} d\xi
  \label{estim_Us_Un}
\end{eqnarray}
We define
\begin{eqnarray}
  \cE_0 &\equiv& \frac{\Delta^2}{V} - \int_{-\hbar\omega_c}^{\hbar\omega_c} \sigma(\xi+\mu) \epsilon_\xi \left(1 - \frac{|\xi|}{\epsilon} \right)\,d\xi + \mu \int_{0}^{\hbar\omega_c}[\sigma(\xi+\mu) - \sigma(-\xi+\mu)] \left( 1 - \frac{\xi}{\epsilon} \right) d\xi
   \label{def_cE0}
\end{eqnarray}
and
\begin{equation}
  E_{qp} \equiv \int_{-\hbar\omega_c}^{\hbar\omega_c}\sigma(\xi+\mu) ( n_{\xi 0} + n_{\xi 1}) \left(\epsilon + \mu\frac{\xi}{\epsilon} \right) d\xi , \label{def_cHpqp}
\end{equation}
such that the total energy of the system, $E = \langle \cH \rangle = E_{0\mu} + \cE_0 + E_{qp}$.
If the DOS is constant, then \cite{PhysicaA.459.276.2016.Anghel}
\begin{eqnarray}
  \cE_0 &=& 2\sum_{k> k_F} (\xi_\bk - \epsilon_\bk) + \frac{\Delta^2}{V}
  = - \frac{\sigma_0\Delta^2}{2} \left[ 1 + 2 \ln\left(\frac{\Delta_0}{\Delta}\right) \right] . \label{estim_Us_Un_sig0}
\end{eqnarray}

We may also write the energy operator in the second quantization,
\begin{equation}
  \hat\cH = E_{0\mu} + \cE_0 + \sum_\bk \left( \epsilon_\bk + \mu \frac{\xi_\bk}{\epsilon_\bk} \right) (\gamma^\dagger_{\bk 0}\gamma_{\bk 0} + \gamma^\dagger_{\bk 1}\gamma_{\bk 1}) . \label{HM_BCS_eff2}
\end{equation}
The BCS gap energy $\Delta$ depends on the quasiparticle populations $n_{\bk i}$ through the self-consistent equation (\ref{def_Delta2}) (or \ref{Eq_int_Delta1}) and $\cE_0$ is given by Eq. (\ref{def_cE0}).
Through the dependence of $\Delta$ on the quasiparticle populations, both $\cE_0$ and $\epsilon_\bk$ depend on the quasiparticle populations.

\section{The variation of the energy with the number of quasiparticle excitations} \label{Var_en_gap}

To calculate the effect of changing the number of quasiparticle excitations on the total energy and total particle number we apply a procedure specific to fractional exclusion statistics, like in Ref. \cite{PhysicaA.459.276.2016.Anghel}.
We divide the space of wave-vectors $\bk$ into elementary volumes, $\delta\bk$, each centered or containing (by definition) the vector $\bk$. Each such volume contains $G_{\delta\bk} \equiv \sigma(\bk)\delta\bk$ states and $N_{\delta\bk} \equiv N_{\delta\bk 0}+N_{\delta\bk 1}$ quasiparticles of types 0 and 1--we denoted by $\sigma(\bk)$ the DOS in the $\bk$ space without summing over the spin projections or quasiparticle types, $0$ and $1$.
Then we can redefine $n_{\bk 0} \equiv N_{\delta\bk 0}/G_{\delta\bk}$ and $n_{\bk 1} \equiv N_{\delta\bk 1} /G_{\delta\bk}$.
In these notations we write the gap energy explicitly as a function of the set of numbers of quasiparticle excitations in the volumes $\delta\bk$, namely $\Delta(\{N_{\delta\bk i}\})$, and Eq. (\ref{def_Delta2}) becomes
\begin{equation}
  1 = \frac{V}{2} \sum_{\delta\bk} \frac{G_{\delta\bk} - N_{\delta\bk 0} - N_{\delta\bk 1}}{\epsilon_\bk} . \label{def_Delta3}
\end{equation}
In Eq. (\ref{def_Delta3}) we assumed that the volumes $\delta\bk$ are sufficiently small, so that all the quasiparticle energies in one such volume may be taken equal to $\epsilon_\bk$.
Then (through $\Delta$) $\cE_0$ and $\epsilon_\bk$'s in the expression of $\cH$ (see Eq. \ref{HM_BCS_eff2}) become functions of the sets of numbers of particles, $\{N_{\delta\bk 0}\}$ and $\{N_{\delta\bk 1}\}$:
\begin{subequations} \label{dcE0_dEqp_dNki}
\begin{eqnarray}
  \frac{\partial \cE_0}{\partial N_{\delta\bk i}} &=& \frac{\partial\cE_0}{\partial\Delta} \frac{\partial\Delta}{\partial N_{\delta\bk i}} \equiv \epsilon_\bk' \label{dcE0_dNki} \\
  %
  \frac{\partial E_{qp}}{\partial N_{\delta\bk i}} &=& \epsilon_\bk + \mu\frac{\xi}{\epsilon} + \frac{\partial\Delta}{\partial N_{\delta\bk i}} \frac{\partial E_{qp}}{\partial\Delta} \equiv \epsilon_\bk + \epsilon_\bk'', \label{dEqp_dNki}
\end{eqnarray}
\end{subequations}
where we used Eqs. (\ref{estim_Us_Un}) and (\ref{def_cHpqp}).
We also notice that $\partial E_{0\mu}/\partial N_{\delta\bk i} = 0$ since $E_{0\mu}$ depends only on the parameter $\mu$ of the model.
The variations with respect to $\Delta$ of $\cE_0$ and $\cE_{qp}$ are
\begin{subequations} \label{dcE_dHqp_dDelta}
\begin{eqnarray}
  \frac{\partial \cE_0}{\partial \Delta} 
  %
  &=&  \frac{2\Delta}{V} - \Delta \int_{-\hbar\omega_c}^{\hbar\omega_c} \sigma(\xi+\mu) \frac{1}{\epsilon} \left(1 - \frac{\mu \xi}{\epsilon^2} \right) \quad {\rm and}  \label{dcE_dDelta} \\
  \frac{\partial E_{qp}}{\partial \Delta} 
  %
  %
  &=& \Delta \int_{-\hbar\omega_c}^{\hbar\omega_c}\sigma(\xi+\mu) \frac{1}{\epsilon}\left(1 - \frac{\mu\xi}{\epsilon^2} \right) d\xi
  - \frac{2 \Delta}{V} 
  + \Delta \mu \int_{-\hbar\omega_c}^{\hbar\omega_c}\sigma(\xi+\mu) ( 1 - n_{\xi 0} - n_{\xi 1}) \frac{\xi}{\epsilon^3} d\xi , \label{dEqp_dDelta}
\end{eqnarray}
\end{subequations}
so from Eqs. (\ref{dcE_dHqp_dDelta}) we obtain
\begin{eqnarray}
  \frac{\partial E}{\partial\Delta} &=& \frac{\partial (\cE_0+E_{qp})}{\partial \Delta} = \Delta \mu \int_{-\hbar\omega_c}^{\hbar\omega_c}\sigma(\xi+\mu) \frac{\xi}{\epsilon^3} d\xi
  - \Delta \mu \int_{-\hbar\omega_c}^{\hbar\omega_c}\sigma(\xi+\mu) ( n_{\xi 0} + n_{\xi 1}) \frac{\xi}{\epsilon^3} d\xi
  . \label{dE_dDelta}
\end{eqnarray}
The first integral in Eq. (\ref{dE_dDelta}) vanishes if $\sigma(\xi) \equiv \sigma_0$ is constant.
The variation of $\Delta$ with $N_{\delta\bl i}$ is calculated from
\begin{eqnarray}
  0 &=& - \Delta\left[ \int_{-\hbar\omega_c}^{\hbar\omega_c} \frac{ \sigma(\xi+\mu)\,d\xi}{(\xi^2+\Delta^2)^{3/2}}
  - \int_{-\hbar\omega_c}^{\hbar\omega_c} \frac{(n_{0 \xi} + n_{1 \xi}) \sigma(\xi+\mu) d\xi}{(\xi^2+\Delta^2)^{3/2}} \right] d\Delta
  - \frac{(d N_{\delta\bl 0} + d N_{\delta\bl 1})}{\sqrt{\xi^2+\Delta^2}} \label{var_Delta_N}
\end{eqnarray}
from where we observe that
\begin{eqnarray}
  \frac{\partial\Delta}{\partial N_{\xi i}} &=& - \left\{\Delta \epsilon \left[ \int_{-\hbar\omega_c}^{\hbar\omega_c} \frac{(1 - n_{0\xi} - n_{1\xi}) \sigma(\xi+\mu) d\xi}{\epsilon^3} \right] \right\}^{-1} . \label{dDelta_dNxi}
\end{eqnarray}
From Eqs. (\ref{dcE0_dEqp_dNki}), (\ref{dcE_dHqp_dDelta}), and (\ref{dDelta_dNxi}) we obtain
\begin{eqnarray}
  \frac{\partial E}{\partial N_{\delta \bk}} &=& \epsilon_\bk + \frac{\mu}{\epsilon} \left\{ \xi
  - \frac{ \int_{-\hbar\omega_c}^{\hbar\omega_c} \sigma(\xi+\mu) ( 1 - n_{\xi 0} - n_{\xi 1}) \frac{\xi d\xi}{\epsilon^3_\xi} }{ \int_{-\hbar\omega_c}^{\hbar\omega_c} (1 - n_{0\xi} - n_{1\xi}) \sigma(\xi+\mu) \frac{d\xi}{\epsilon^3_\xi} } \right\}
  \equiv \epsilon_\bk + \delta \epsilon_\bk \equiv \tilde \epsilon_\bk \label{dE_dNbk_tot}
\end{eqnarray}
From Eq. (\ref{dE_dNbk_tot}) we observe that the quasiparticle energy is $\tilde \epsilon_\bk$ instead of $\epsilon_\bk$.

\section{The variation of the number of particles with the number of quasiparticle excitations} \label{Var_num_part}

From Eq. (\ref{N_Nmu_sum}) or Eq.  (\ref{N_Nmu_int}) we obtain
\begin{subequations} \label{dN_dDelta_dNbk}
\begin{eqnarray}
  \frac{\partial N}{\partial \Delta} &=& \Delta \int_{-\hbar\omega_c}^{\hbar\omega_c}\sigma(\xi+\mu) \left( 1 - n_{\xi 0} - n_{\xi 1} \right) \frac{\xi d\xi}{\epsilon^3} \quad {\rm and} \label{dN_dDelta} \\
  \frac{\partial N}{\partial N_{\bk i}} &=& \frac{\xi}{\epsilon} ,
  \label{dN_dNbk}
\end{eqnarray}
\end{subequations}
from where it follows
\begin{eqnarray}
  \frac{d N}{d N_{\xi i}} &=& \frac{1}{\epsilon} \left\{ \xi - \frac{ \int_{-\hbar\omega_c}^{\hbar\omega_c}\sigma(\xi+\mu) \left( 1 - n_{\xi 0} - n_{\xi 1} \right) \frac{\xi d\xi}{\epsilon^3_\xi} }{ \int_{-\hbar\omega_c}^{\hbar\omega_c} \sigma(\xi+\mu) (1 - n_{0\xi} - n_{1\xi}) \frac{d\xi}{\epsilon^3_\xi} } \right\} . \label{dN_dNbk_tot}
\end{eqnarray}

\section{The equilibrium particle distribution} \label{sec_Eq_ples} 

Assuming that the system is in contact with a heat and particle reservoir of temperature $T$ and chemical potential $\mu_R$ (eventually $\mu_R \ne \mu$), using Eqs. (\ref{N_Nmu_int_sconst2}) and (\ref{HM_BCS_eff2}) we calculate the grandcanonical partition function
\begin{eqnarray}
  \ln(\cZ)_{\beta\mu} &=& - \sum_{\bk i} [(1 - n_{\bk i}) \ln(1 - n_{\bk i}) + n_{\bk i} \ln n_{\bk i} ] - \beta (E-\mu_R N) . \label{cZ_beta_mu}
\end{eqnarray}
To find the equilibrium populations, we maximize $\ln(\cZ)_{\beta\mu}$ with respect to the populations $n_{\bk i}$ or (equivalently) to the numbers of particles in the grains $N_{\delta\bk i}$.
For this, we calculate first
\begin{eqnarray}
  E-\mu_R N &=& E_{0\mu} + \cE_0 + E_{qp} - \mu_R (N-N_\mu) - \mu_R N_\mu \nonumber \\
  &=& E_{0\mu} - \mu_R N_\mu + \frac{\Delta^2}{V} \nonumber \\
  %
  && - \int_{-\hbar\omega_c}^{0}\sigma(\xi+\mu) [\epsilon + (\mu - \mu_R)] \left( 1 - \frac{|\xi|}{\epsilon} \right) d\xi
  - \int_{0}^{\hbar\omega_c}\sigma(\xi+\mu) [\epsilon - (\mu-\mu_R)] \left( 1 - \frac{\xi}{\epsilon} \right) d\xi \nonumber \\
  && + \int_{-\hbar\omega_c}^{\hbar\omega_c}\sigma(\xi+\mu) ( n_{\xi 0} + n_{\xi 1}) \left[\epsilon + (\mu-\mu_R)\frac{\xi}{\epsilon} \right] d\xi \label{Em_muN}
\end{eqnarray}
From Eqs. (\ref{cZ_beta_mu}) and (\ref{Em_muN}) we obtain, using Eqs. (\ref{dE_dNbk_tot}) and (\ref{dN_dNbk_tot}),
\begin{equation}
  \frac{\partial\ln(\cZ)_{\beta\mu}}{\partial n_{\bk i}} = \ln\frac{1-n_{\bk i}}{n_{\bk i}} - \beta \left\{\epsilon_\bk - \frac{\mu_R - \mu}{\epsilon_\bk} \left[ \xi_\bk - \frac{ \int_{-\hbar\omega_c}^{\hbar\omega_c} \sigma(\xi+\mu) \frac{\xi}{\epsilon^3} \, d\xi
  - \int_{-\hbar\omega_c}^{\hbar\omega_c} \sigma(\xi+\mu) ( n_{\xi 0} + n_{\xi 1} ) \frac{\xi}{\epsilon^3} \, d\xi}
  { \int_{-\hbar\omega_c}^{\hbar\omega_c} \frac{(1 - n_{0\xi} - n_{1\xi}) \sigma(\xi+\mu) d\xi}{\epsilon^3} } \right]\right\} = 0, \label{dlnZ_dnki}
\end{equation}
Equation (\ref{dlnZ_dnki}) leads to the Fermi populations
\begin{equation}
  n_{\bk i} = \frac{1}{e^{\beta(\epsilon_{\bk}-\tilde\mu)}+1} , \label{pop_til_eps}
\end{equation}
with an effective chemical potential dependent on the energy,
\begin{equation}
  \tilde\mu \equiv \frac{\mu_R - \mu}{\epsilon_\bk} \left[ \xi_\bk - \frac{ \int_{-\hbar\omega_c}^{\hbar\omega_c} \sigma(\xi+\mu) ( 1 - n_{\xi 0} - n_{\xi 1} ) \frac{\xi}{\epsilon^3} \, d\xi }
  { \int_{-\hbar\omega_c}^{\hbar\omega_c} \frac{(1 - n_{0\xi} - n_{1\xi}) \sigma(\xi+\mu) d\xi}{\epsilon^3} } \right] . \label{def_tilde_mu}
\end{equation}
If $\mu = \mu_R$, we recover the standard BCS formalism, as we did in Ref. \cite{PhysicaA.459.276.2016.Anghel}.


The calculations simplify considerably if we assume a constant DOS $\sigma(\epsilon^{(0)}) \equiv \sigma_0$. Then, using again $n_{\xi 0} = n_{\xi 1} \equiv n_\xi$, Eq. (\ref{Em_muN}) becomes
\begin{eqnarray}
  E-\mu_R N &=& E_{0\mu} - \mu_R N_\mu + \frac{\Delta^2}{V}
  - \sigma_0 \Delta^2 \ln \left( \frac{2\hbar\omega_c}{\Delta} \right)
  + 2 \sigma_0  \int_{-\hbar\omega_c}^{\hbar\omega_c} n_\xi \left[\epsilon + (\mu-\mu_R)\frac{\xi}{\epsilon} \right] d\xi
  \label{Em_muN_s0}
\end{eqnarray}
whereas Eq. (\ref{def_tilde_mu}) becomes
\begin{equation}
  \tilde\mu \equiv \frac{\mu_R - \mu}{\epsilon_\bk} \left[ \xi_\bk - \frac{ \int_{-\hbar\omega_c}^{\hbar\omega_c} ( 1 - n_{\xi 0} - n_{\xi 1} ) \frac{\xi}{\epsilon^3} \, d\xi }
  { \int_{-\hbar\omega_c}^{\hbar\omega_c} \frac{(1 - n_{0\xi} - n_{1\xi}) d\xi}{\epsilon^3} } \right]
  \equiv \frac{\mu_R - \mu}{\epsilon_\bk} \left[ \xi_\bk - F \right] .
  \label{def_tilde_mu_sigma0}
\end{equation}
Equation (\ref{def_tilde_mu_sigma0}), together with the expression for $n_{\bk i}$ (\ref{pop_til_eps}),
\begin{subequations}\label{def_F_sigma0_set}
\begin{eqnarray}
  F &\equiv& \frac{ \int_{-\hbar\omega_c}^{\hbar\omega_c} ( 1 - n_{\xi 0} - n_{\xi 1} ) \frac{\xi}{\epsilon^3} \, d\xi }
  { \int_{-\hbar\omega_c}^{\hbar\omega_c} \frac{(1 - n_{\xi 0} - n_{\xi 1}) d\xi}{\epsilon^3} } , \label{def_F_sigma0} \\
  n_{\xi i} &=& \frac{1}{e^{\beta[\epsilon_{\xi}-(\mu_R-\mu)(\xi - F)/\epsilon_\xi]}+1} , \label{pop_til_eps_sigma0} 
\end{eqnarray}
\end{subequations}
plus the equation for $\Delta$ (\ref{Eq_int_Delta1}) form the self-consistent set of equations.
Introducing the dimensionless variable $x_F \equiv \beta F$ and changing the variables in the integrals of Eqs. (\ref{def_F_sigma0_set}) and (\ref{Eq_int_Delta1}) this self-consistent set of equations can be written
\begin{subequations}\label{def_xF_sigma0_set}
\begin{eqnarray}
  x_F &=& \frac{\int_{y}^{\beta\hbar\omega_c} \frac{( n_{-x} - n_{x} )\, dx}{x^2}}{\int_{y}^{\beta\hbar\omega_c} \frac{(1 - n_{-x} - n_{x})\, dx}{x^2 \sqrt{x^2-y^2}}} , \label{def_xF_sigma0} \\
  n_x &=& \frac{1}{e^{x-y_R\left(\sqrt{x^2-y^2} - x_F\right)/x}+1} , \label{pop_til_eps_sigma0_p} \\
  n_{-x} &=& \frac{1}{e^{x-y_R\left(-\sqrt{x^2-y^2} - x_F\right)/x}+1} \label{pop_til_eps_sigma0_n}
\end{eqnarray}
where $x\equiv \beta\epsilon$, $y \equiv \beta\Delta$, and $y_R \equiv \beta (\mu_R - \mu)$.
We also wrote explicitly the populations for the positive and negative branches, $\xi = \sqrt{\epsilon^2 - \Delta^2}$ (\ref{pop_til_eps_sigma0_p}) and $\xi = - \sqrt{\epsilon^2 - \Delta^2}$ (\ref{pop_til_eps_sigma0_n}), respectively.
Rewriting Eq. (\ref{Eq_int_Delta1}) for constant DOS in these notations for computational convenience, we obtain
\begin{equation}
  \frac{1}{\sigma_0 V} 
  = \int_y^{\beta\hbar\omega_c} \frac{1 - n_{-x} - n_{x}}{\sqrt{x^2 - y^2}} dx 
  \label{Eq_int_Delta1_2}
\end{equation}
\end{subequations}
We observe that Eqs. 
(\ref{def_xF_sigma0_set}) are symmetric under the exchange $y_R \to -y_R$, $x_F \to -x_F$, and $\xi \to -\xi$.
Solving self-consistently the set of equations (\ref{def_xF_sigma0_set}) we obtain the equilibrium populations.

\begin{figure}[t]
  \centering
  \includegraphics[width=7cm,keepaspectratio=true]{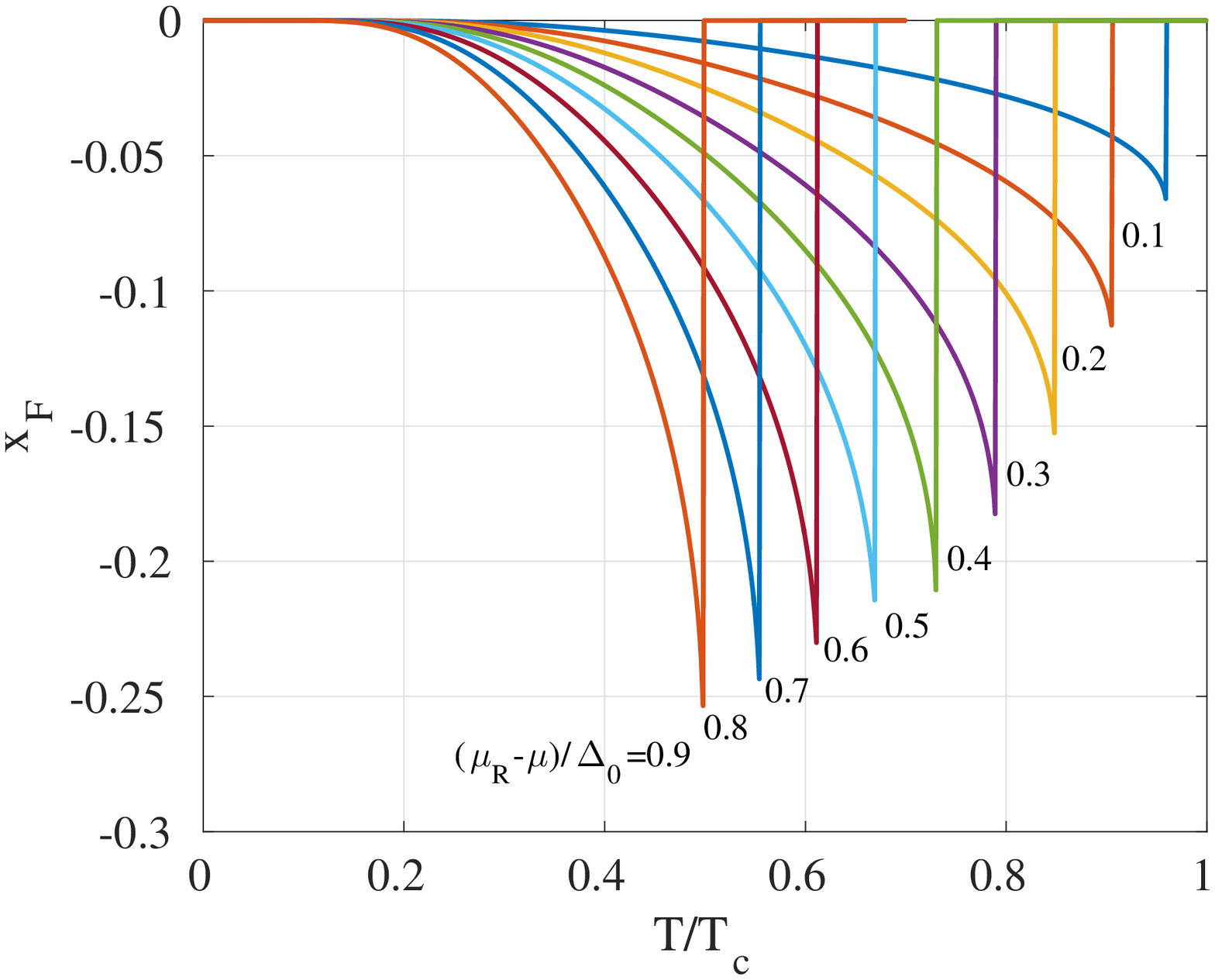}
  \includegraphics[width=7cm,keepaspectratio=true]{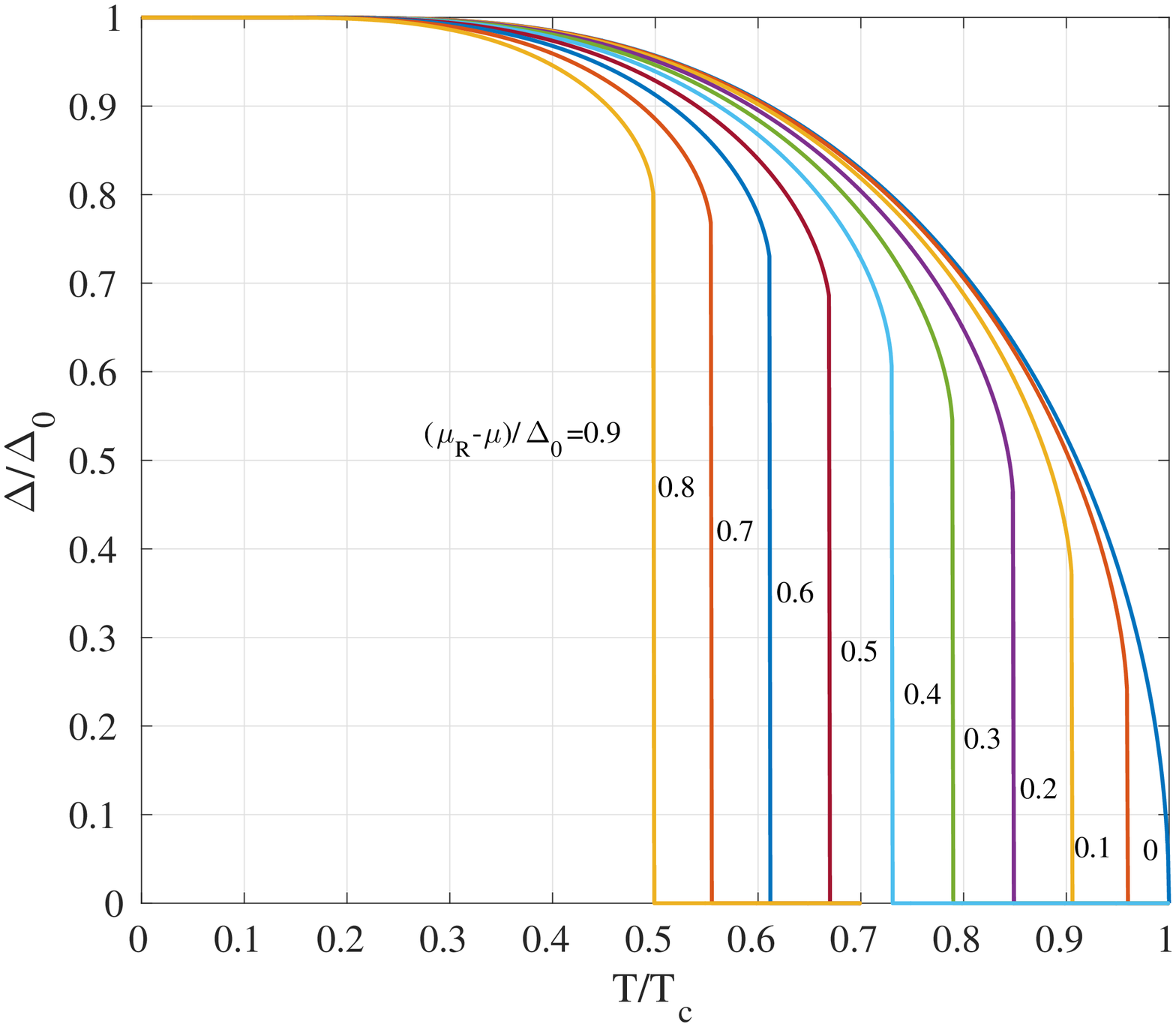}
  \caption{(Color online) The solutions $x_F$ and $\Delta$ of Eqs. (\ref{def_xF_sigma0_set}) for $(\mu_R-\mu)/\Delta_0 = 0, 0.1, \ldots, 0.9$; $\Delta_0$ is the value of the gap energy at $T=0$ and $T_c$ is the BCS critical temperature.}
  \label{Delta_xF_vs_T}
\end{figure}

\begin{figure}[t]
  \centering
  \includegraphics[width=8 cm,bb=0 0 591 455,keepaspectratio=true]{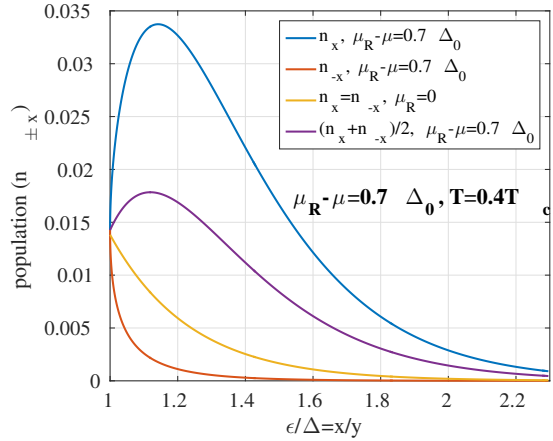}
  \caption{(Color online) The branch imbalance, $n_x \ne n_{-x}$, for $\mu_R - \mu = 0.7 \Delta_0$. For comparison we show also the typical BCS population, $n_x = n_{-x}$, for $\mu_R = \mu$, and the average $(n_x + n_{-x})/2$, for $\mu_R - \mu = 0.7 \Delta_0$.}
  \label{pops_mu0o7Delta0_T0o4Tc}
\end{figure}

In Fig. \ref{Delta_xF_vs_T} we plot the solutions of Eqs. (\ref{def_xF_sigma0_set}) for $(\mu_R-\mu)/ \Delta_0 = 0, 0.1, \ldots, 0.9$, where $\Delta_0$ is the value of the gap energy at $T=0$,
corresponding to the standard BCS value.
If $\mu_R = \mu$, $x_F = 0$, and we recover the BCS results.
We notice that if $\mu_R \ne \mu$, the gap energy is always smaller than the BCS gap and, at a temperature smaller than the BCS critical temperature $T_c$, the superconductor has a first order phase transition to the normal state.
We also observe that if $y_R\ne 0$, a branch imbalance appears: the populations of the branches with $\xi>0$ (\ref{pop_til_eps_sigma0_p}) and $\xi<0$ (\ref{pop_til_eps_sigma0_n}) are not equal. In Fig. \ref{pops_mu0o7Delta0_T0o4Tc} we plot the populations for $\xi > 0$ (blue line) and $\xi<0$ (red line) for $\mu_R-\mu = 0.7\Delta_0$.
For comparison we plot also the Fermi populations for $\mu_R = \mu$ (yellow line) and the average $(n_x + n_{-x})/2$ for $\mu_R - \mu = 0.7\Delta_0$.
The fact that the population $n_x$ increases with $x$ for $x/y = \epsilon/\Delta$ close to 1 (blue line) is due to the fact that $\epsilon$ is not the quasiparticle energy of the system in the Landau's sense (i.e. $\epsilon_\bk \ne \partial E/ \partial n_\bk$, see Eq. \ref{dE_dNbk_tot}).

\subsection{Low temperature limit for constant DOS} \label{subsub_lowT}

To analyze the equilibrium populations in the low temperature limit, we start from Eqs. (\ref{def_xF_sigma0_set}). Let us discuss the case $y_R > 0$ since the case $y_R < 0$ can be recovered from this by the replacement $x_F \to - x_F$ and exchanging $n_\xi$ with $n_{-\xi}$. First, we observe that if $y_R/y < 1$ (i.e. $\mu_R - \mu < \Delta$) we always have a solution with $\lim_{T\to 0} x_F = 0$ which corresponds to $n_{\xi} = n_{-\xi} = 0$ for any $\xi \in (-\hbar\omega_c, \hbar \omega_c)$. From Eq. (\ref{Eq_int_Delta1_2}), with $n_{\xi} = n_{-\xi} = 0$, we obtain an energy gap at zero temperature $\Delta(T=0) = \Delta_0$.

In the more general case, let us analyze the argument of the exponential function in the denominator of $n_x$ and $n_{-x}$ of Eqs. (\ref{def_xF_sigma0_set}).
If we write $n_x \equiv \left\{ \exp[\beta m_x] + 1 \right\}^{-1}$ and $n_{-x} \equiv \left\{ \exp[\beta m_{-x}] + 1 \right\}^{-1}$, then
\begin{subequations} \label{defs_mx}
\begin{eqnarray}
  m_x &\equiv& \Delta \left[ r - \frac{a}{r} \left( \sqrt{r^2 - 1} - b \right) \right] , \label{def_mx} \\
  m_{-x} &\equiv& \Delta \left[ r - \frac{a}{r} \left( - \sqrt{r^2 - 1} - b \right) \right] , \label{def_mmx}
\end{eqnarray}
\end{subequations}
where $r = \epsilon/\Delta = x/y$, $a = (\mu_R - \mu)/\Delta = y_R/y$, and $b = F/\Delta = x_F/y$. To see if we have a solution with $x_F = 0$, we set $b=0$ in Eq. (\ref{def_mx}) and we observe that $m_x \ge 0$ for any $y_r \le 2y$.
But if $m_x \ge 0$, then also $m_{-x}\ge 0$ and therefore $\lim_{T\to0} \beta m_x = \lim_{T\to0} \beta m_{-x} = \infty$. This implies that in the limit $T=0$, $n_x = n_{-x} = 0$ and therefore $\lim_{T\to 0} \Delta(T) = \Delta_0$.
In other words, for $|\mu_R - \mu| \le 2\Delta_0$, the energy gap at $T = 0$ is $\Delta_0$, but the transition temperature decreases with increasing $\mu_R - \mu$.

If $\mu_R-\mu > 2\Delta_0$, then $a > 2$. In such a case $m_x$ may take both, positive and negative values, whereas $m_{-x}$ is always positive.
In the limit $T\to 0$, Eqs. (\ref{def_xF_sigma0_set}) cannot have solutions corresponding to $b = 0$ (i.e. $x_F = 0$) and therefore a quasiparticle imbalance exists even at zero temperature.

\section{Conclusions} \label{discussion}


If the Cooper pairing potential acts between particles with kinetic energy (free single-particle energies) in an interval centered at $\mu$ and the absolute value of the difference between $\mu$ and the chemical potential of the system $\mu_R$ is smaller or equal to $2\Delta_0$ ($\Delta_0$ being the standard BCS gap energy at zero temperature), the superconducting phase appears, with a gap energy at zero temperature equal to $\Delta_0$.
The superconductor-normal metal phase transition temperature is highest if $\mu = \mu_R$ and decreases to zero as $|\mu_R - \mu|$ increases to $2\Delta_0$.
The phase transition is of the first order if $\mu_R \ne \mu$ and is of the second order only if $\mu_R = \mu$.
The system of equations that give the gap energy and the populations (Eqs. \ref{def_xF_sigma0_set}) may have more than one solution at the same temperature and if $\mu_R \ne \mu$ a branch imbalance appears, although the system is at equilibrium.
In previous studies of branch imbalance (see for example Refs. \cite{PhysRevLett.28.1363.1972.Clarke, PhysRevLett.28.1366.1972.Tinkham, PhysRevB.6.1747.1972.Tinkham, PhysRevB.92.144506.2015.Miller, PhysRevB.93.220501.2016.Quay}), the systems were out of equilibrium and the imbalance was described by attributing different chemical potentials to the quasiparticles and to the BCS condensate of pairs.
The analysis of such systems will constitute a separate study, but we stress the fact that here we describe the system unitarily, with a unique chemical potential.
The variation of this chemical potential along the superconductor may describe a non-equilibrium situation as well.

Although there is no clear reason why the energetic interval in which the BCS pairing interaction should be centered at the Fermi level, these aspects of the BCS theory have not been investigated before, to the best of our knowledge. Physically, the Fermi level (or the Fermi surface in the $\bk$ space) can be changed by doping or by applied pressure and $\mu_R$ can be moved away (or towards) $\mu$. Our results indicate that in such a situation even the standard BCS theory for isotropic solids implies the formation of a kind of a \textit{superconducting dome}. This model can be refined further, by adding other effects, like the anisotropy of the system.
Nevertheless, based on these results we consider that the full and correct picture of the superconducting phase transition cannot be grasped unless we dissociate the range of pairing interaction in the $\bk$ space from the Fermi surface.

\section{Acknowledgments}

Discussions with Dr. Al. Yu. Cherny, Dr. Yu. Shukrinov, Dr. S. Cojocaru, Prof. Yu. M. Galperin and Prof. J. Bergli are gratefully acknowledged.
This work has been financially supported by CNCSIS-UEFISCDI (project IDEI 114/2011) and ANCS (project PN-09370102 PN 16420101/ 2016). Travel support from Romania-JINR Collaboration grants 4436-3-2015/2017, 4342-3-2014/2015, and the Titeica-Markov program is gratefully acknowledged.

\section{Author contributions}

DVA introduced the idea, formulated the concept, wrote the first draft of the manuscript, and participated in the interpretation of the results. He is also responsible for the analytical and numerical calculations. GAN contributed to the writing of the manuscript and to the interpretation of the results. He also did the numerical calculations independently, to compare the results.


\end{document}